\documentclass[twocolumn,10pt]{article} 

\usepackage[square,numbers,sort&compress,comma]{natbib}

\usepackage{amsmath}
\usepackage{amssymb}
\usepackage{caption}
\usepackage{graphicx}
\usepackage{latexsym}
\usepackage{times}
\usepackage[pagewise]{lineno}
\usepackage{hyperref}
\usepackage{algorithm}
\usepackage{algorithmic,color}

\topmargin - 12pt 
\renewenvironment{abstract}%
              {
               \small
               {\bfseries \abstractname}
               \par
               \vspace{10pt}
              }

\renewcommand\abstractname{Abstract}

\newcommand{\nomenclature}
              [1]
              {
               \bgroup
               \flushleft
               \small\bf
               #1
               \par
               \egroup
              }

\renewcommand{\section}
              [1]
              {
               \bgroup
               \flushleft
               \small\bf
               \refstepcounter{section}
               \arabic{section}. #1
               \par
               \egroup
              }

\renewcommand{\subsection}
              [1]
              {
               \bgroup
               \flushleft
               \small\em
               \refstepcounter{subsection}
               \arabic{section}.
               \arabic{subsection}. #1
               \par
               \egroup
              }

\renewcommand{\subsubsection}
              [1]
              {
               \bgroup
               \flushleft
               \small\em
               \refstepcounter{subsubsection}
               \arabic{section}.
               \arabic{subsection}.
               \arabic{subsubsection}. #1
               \par
               \egroup
              }

  \newcommand{\acknowledgement}
              [1]
              {
               \bgroup
               \flushleft
               \small\bf
               #1
               \par
               \egroup
              }

  \newcommand{\sectionbib}
              [1]
              {
               \bgroup
               \flushleft
               \small\bf
               #1
               \par
               \egroup
              }

\begin{document}

\title{\LARGE An integrated framework for accelerating reactive flow simulation using GPU and machine learning models}

\author{{\large Runze Mao$^{a,b}$, Yingrui Wang$^{c}$, Min Zhang$^{a,b}$, Han Li$^{a,b}$,}\\[10pt]{\large  Jiayang Xu$^{b}$, Xinyu Dong$^{a,b}$, Yan Zhang$^{d,e}$, Zhi X. Chen$^{a,b,*}$}\\[10pt]
        {\footnotesize \em $^a$State Key Laboratory of Turbulence and Complex Systems, Aeronautics and Astronautics, College of Engineering,}\\[-5pt]
        {\footnotesize \em Peking University, Beijing, 100871, China}\\[-5pt]
        {\footnotesize \em $^b$AI for Science Institute (AISI), Beijing, 100080, China}\\[-5pt]
        {\footnotesize \em $^c$Shanghai SenseTime Intelligent Technology Co., LTD, Shanghai, 200233, China}\\[-5pt]
        {\footnotesize \em $^d$CAEP Software Center for High Performance Numerical Simulation, Beijing 100088, China}\\[-5pt]
        {\footnotesize \em $^e$Institute of Applied Physics and Computational Mathematics, Beijing 100088, China}}

\date{}


\small
\baselineskip 10pt


\twocolumn[\begin{@twocolumnfalse}
\vspace{50pt}
\maketitle
\vspace{40pt}
\rule{\textwidth}{0.5pt}
\begin{abstract} 
Recent progress in artificial intelligence (AI) and high-performance computing (HPC) have brought potentially game-changing opportunities in accelerating reactive flow simulations. In this study, we introduce an open-source computational fluid dynamics (CFD) framework that integrates the strengths of machine learning (ML) and graphics processing unit (GPU) to demonstrate their combined capability. Within this framework, all computational operations are solely executed on GPU, including ML-accelerated chemistry integration, fully-implicit solving of PDEs, and computation of thermal and transport properties, thereby eliminating the CPU-GPU memory copy overhead. 
Optimisations both within the kernel functions and during the kernel launch process are conducted to enhance computational performance. 
Strategies such as static data reorganisation and dynamic data allocation are adopted to reduce the GPU memory footprint. The computational performance is evaluated in two turbulent flame benchmarks using quasi-DNS and LES modelling, respectively. Remarkably, while maintaining a similar level of accuracy to the conventional CPU/CVODE-based solver, the GPU/ML-accelerated approach shows an overall speedup of over two orders of magnitude for both cases. This result highlights that high-fidelity turbulent combustion simulation with finite-rate chemistry that requires normally hundreds of CPUs can now be performed on portable devices such as laptops with a medium-end GPU.

\end{abstract}
\vspace{10pt}
\parbox{1.0\textwidth}{\footnotesize {\em Keywords:} Compressible reacting flow; GPU acceleration; HPC; machine learning; Chemical kinetics; LES; Quasi-DNS}
\rule{\textwidth}{0.5pt}
\vspace{10pt}

*Corresponding author.\\
\textit{E-mail address:} chenzhi@pku.edu.cn (Zhi X. Chen).

\end{@twocolumnfalse}] 

\clearpage

\section{Introduction\label{sec:introduction}} \addvspace{10pt}

The imperative to decarbonise combustion systems has escalated due to the pressing need to mitigate climate change issues. 
To further advance reacting flow systems in the energy transition era, the development of high-fidelity numerical tools has become a central topic \cite{RN53}. However, modelling methods that involve the direct interplay of turbulent flow and detailed chemistry still pose formidable costs. The majority of the computational expense is attributed to stiff chemistry and multi-species transport \cite{RN108}, which require game-changing acceleration techniques, at both software and hardware levels.

Recent developments in artificial intelligence (AI) and high-performance computing (HPC) have brought inspirations for the entire scientific computing community \cite{wang2023scientific}. Machine learning (ML) have emerged as a new, efficient, powerful and easy-to-use modelling paradigm. On the hardware side, the graphics processing units (GPUs) have become a driving force for the development of computing infrastructure \cite{top500}. For combustion research, prior efforts attempting to accommodate these new techniques mainly focused on accelerating the most time-consuming chemical reaction rate integration. Leveraging ML algorithms, many groups have attempted neural networks to predict chemical kinetics \cite{WAN2020119, zhang2022multi, ding2021machine}. From a hardware perspective, it is also natural to accelerate the ordinary differential equations (ODEs) using GPUs, which offer enhanced performance for these communication-free computations \cite{spafford2010accelerating, RN106, barwey2021neural, niemeyer2017pyjac}.

To facilitate the integration of ML and HPC in reactive flow simulation, we recently developed an open-source platform {DeepFlame}\footnote{\href{https://github.com/deepmodeling/deepflame-dev}{https://github.com/deepmodeling/deepflame-dev}}, based on a code coupling of OpenFOAM, Cantera, and Torch libraries. Using DeepFlame \cite{mao2023deepflame}, we demonstrated that when the chemical source term is computed through the inference of deep neural networks (DNNs, developed in \cite{zhang2022multi}) on GPU, as compared to the conventional iterative approach of solving stiff chemical ODEs, a significant speedup of two orders of magnitude was achieved, even just for a small mechanism with 9 species and 12 reactions. 

However, currently not all computational procedures in reactive CFD can be reliably accelerated by ML-driven models. The flow and species transport is governed by a system of partial differential equations (PDEs). Novel data-driven methods such as physics informed neural network (PINN) \cite{raissi2019physics} seems promising but yet to mature for complex geometry and reactive flows. In the pursuit of speeding up the PDE computations, HPC hardware accelerators, particularly GPUs, have received significant attention within the combustion community \cite{RN41,levesque2012hybridizing,uranakara2023accelerating,bielawski2023highly}. This can be mainly attributed to two key factors: (i) GPU is proven to be advantageous for CFD due to its inherent data parallelism in calculations, i.e. the per-cell parallelism in finite volume method (FVM) \cite{RN108}; and (ii) GPU features a greater number of computing units with simplified control mechanisms within each chip, affording powerful computing capacity and high throughput.

Le et al. \cite{RN41} pioneered the development of one of the first reactive-flow GPU solvers. 
Their work demonstrates a maximum speedup of 40X compared to a single CPU thread. Levesque et al. \cite{levesque2012hybridizing} successfully ported Sandia National Laboratory's Direct Numerical Simulation (DNS) code S3D to GPUs, utilising OpenACC directives in conjunction with MPI and OpenMP to achieve an architecture-agnostic, multi-level parallelism. Recently, Uranakara et al. \cite{uranakara2023accelerating} achieved a maximum speedup of 7X by integrating the GPU-based chemical solver UMChemGPU \cite{barwey2021neural} into the GPU-compatible DNS solver KARFS. Bielawski et al. \cite{bielawski2023highly} developed a high-speed reacting flow GPU solver based on OpenFOAM, , exhibiting an impressive parallel scalability extending to thousands of GPUs.

Despite these notable advances, several challenges persist. First, existing studies have focused on explicit methods for flow transport terms. It is well-known that implicit FVM is of practical importance for low Mach-numbers flows since the computational efficiency is not constrained by the Courant number limit. However, implicit methods require additional matrix discretisation and solving linear equations, both are non-trivial for GPU implementation. Second, an integrated framework to accommodate both ML models and GPU-like accelerator hardware is still absent and their combined performance is yet be demonstrated. 

In this work, we present a GPU porting and performance study of a fully implicit solver called~{\em dfLowMachFoam} in the {DeepFlame} open-source framework~\cite{mao2023deepflame}. The implementation, based on CUDA kernel functions, closely mirrors OpenFOAM, including the discretisation of conservation equations and the assembly of sparse linear matrices for implicit solving. The GPU-accelerated sparse linear solver, AmgX by NVIDIA \cite{RN60}, is employed to solve the implicit linear system. In addition, various optimisations are implemented to enhance computational performance and reduce GPU memory footprint. Chemistry integration is performed via inferencing pretrained DNN models. Two case studies are presented: (i) quasi-DNS of a turbulent H$_2$-air diffusion flame and, (ii) large eddy simulation (LES) of a lab-scale turbulent stratified CH$_4$-air premixed flame. 
Significant acceleration over two orders of magnitude is achieved for both cases using GPU with DNN and detailed discussion is provided. 

The remainder of this paper is organised as follows. In Section~2, we present the theoretical models governing fluid dynamics and chemistry in this study. Section~3 introduces the implementation of the GPU-based solver and outlines the primary optimisations undertaken. Two validation cases, along with a presentation of computational performance, are detailed in Section~4. Finally, the conclusions are summarised in Section~5.

\section{Modeling methodology} \addvspace{10pt}

The solver introduced here uses a fully implicit pressure-based solving procedure for low-speed reactive flows with the pressure-implicit split-operator (PISO) algorithm. 
As per OpenFOAM's standard numeric, the solver employs a cell-centred finite volume scheme with second-order accuracy for discretising the conservation equations and up to fourth-order-accurate for the face flux interpolation \cite{RN111}. Detailed chemical kinetics and molecular transport models are implemented via the Cantera interface. When the mesh resolves the smallest flow and chemical scales, the modelling fidelity at the quasi-DNS level can be achieved. For the LES case presented later, subgrid scale (SGS) closure models are required. The Smagorinsky model is used for the SGS turbulence, and the partially-stirred reactor (PaSR) model \cite{evans2019generalisation} is chosen to account for the SGS turbulence-chemistry interaction. 

The chemical reaction rate integration can be performed using two options: CVODE \cite{10.1145/1089014.1089020} on CPU or DNN on GPU. Detailed validation for the accuracy and generalisation ability of the DNN method is available in \cite{mao2023deepflame}. In this study, we employed individual neural networks to predict the evolution of each component, excluding inert gases. The network features three hidden layers comprising 1600, 800, and 400 perceptrons, respectively. The input layer parameters include temperature, pressure and mass fractions, while the output is the rate of change for a given species. Uniform across all networks, the Gaussian Error Linear Unit (GELU) is chosen as the activation function, and the Adam algorithm is used for hyper-parameter optimisation.

\section{Implementation and optimisations} \addvspace{10pt}

\subsection{GPU implementation} \addvspace{10pt}

\begin{figure*}[ht]
\centering
\includegraphics[width=110mm]{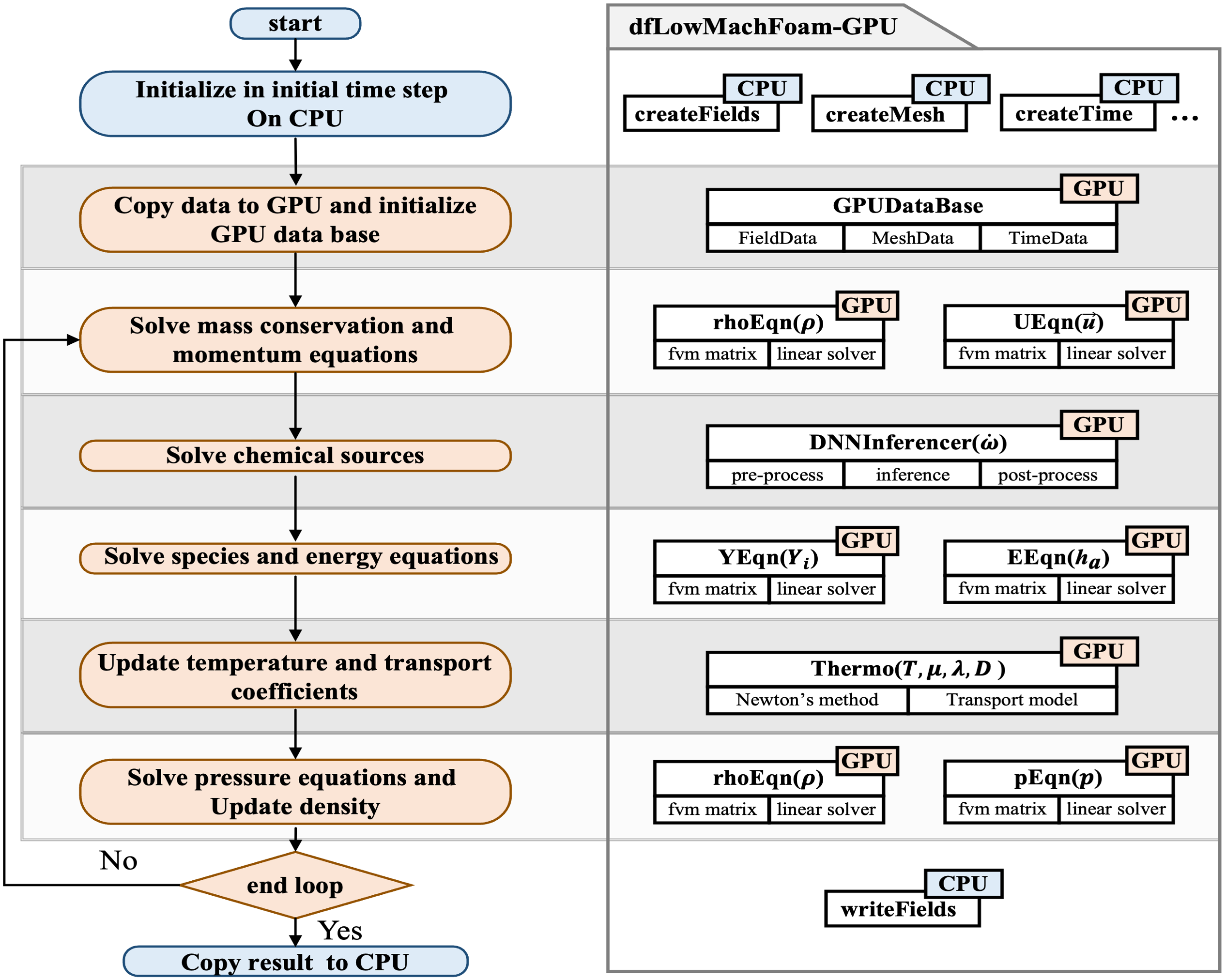}
\caption{Computational Workflow and Code Structure of GPU-based solver {\em dfLowMachFoam}.}
\label{workflow}
\end{figure*}

In this work, the primary objective of GPU porting is to migrate a significant portion of computational operations to the GPU while minimising data transfer between the GPU and CPU. To realise this objective, the internal computational workflow and code structure of the GPU-based solver are meticulously designed, as illustrated in Fig.~\ref{workflow}. Operations highlighted in orange denote computations conducted on the GPU, while those in blue signify CPU-based processes. The functions within the shadowed box aligned with the specified computation process represent the corresponding sections in the code. 

As illustrated in Fig.~\ref{workflow}, it is important to note that the CPU is exclusively engaged in initialising data at the initial time step and writing simulation results at the conclusion. Once the initial data is transferred to the GPU, all the computation required for time advancement are solely performed there. Basically, the operations on GPU can be categorised into three typical types: i) Solving the PDEs: Five distinct PDEs are encompassed in the solver for resolving conservation equations related to mass, momentum, species, energy, and pressure. We employ specialised classes to tackle each PDE. 
ii) Solving the chemical source. As introduced in Section~2, DNN method is adopted here for chemistry integration. The related operations are encapsulated in class {\em DNNInference} and implemented based on libTorch. iii) Calculating the thermophysical and transport property. These operations involve Newton's method and high-order temperature polynomials, conducted by the class {\em Thermo} with CUDA functions. 

\begin{algorithm}[t]\footnotesize
    \caption{implicit discretization of Laplacian term}
    \begin{algorithmic}[1]
    \label{laplacian}
        \STATE \textless\textbf{GPU kernel begin}\textgreater
        \COMMENT{internal field}
        \STATE faceid$\gets$threadIdx.x+blockIdx.x$\times$blockDim.x
        \STATE $\gamma_f$ $\gets$ interpolation($w, \gamma_c$) \label{algo1:l3}
        \STATE $upper$, $lower$ $\gets$ $\gamma_f \delta_f S_f$ \label{algo1:l4}
        \STATE \textbf{atomicAdd}($diag$[$ownCellIndex$[faceid]], $-upper$[faceid]) \label{algo1:l5}
        \STATE \textbf{atomicAdd}($diag$[$neighborCellIndex$[faceid]], $-lower$[faceid]) \label{algo1:l6}
        \STATE \textless\textbf{GPU kernel end}\textgreater
        \FOR{int bouid=1 to $N_{bou}$} \label{algo1:l8}
        \STATE \textless\textbf{GPU kernel begin}\textgreater
        \COMMENT{boundary field}
        \STATE $internalCoeffs, boundaryCoeffs$ $\gets$ \\ updateBouCoeffs($bouType, \gamma_{bou}, S_{bou}$)
        \STATE \textless\textbf{GPU kernel end}\textgreater
        \ENDFOR \label{algo1:l12}
    \end{algorithmic}
\end{algorithm}

\begin{figure}[h!]
\centering
\includegraphics[width=172pt]{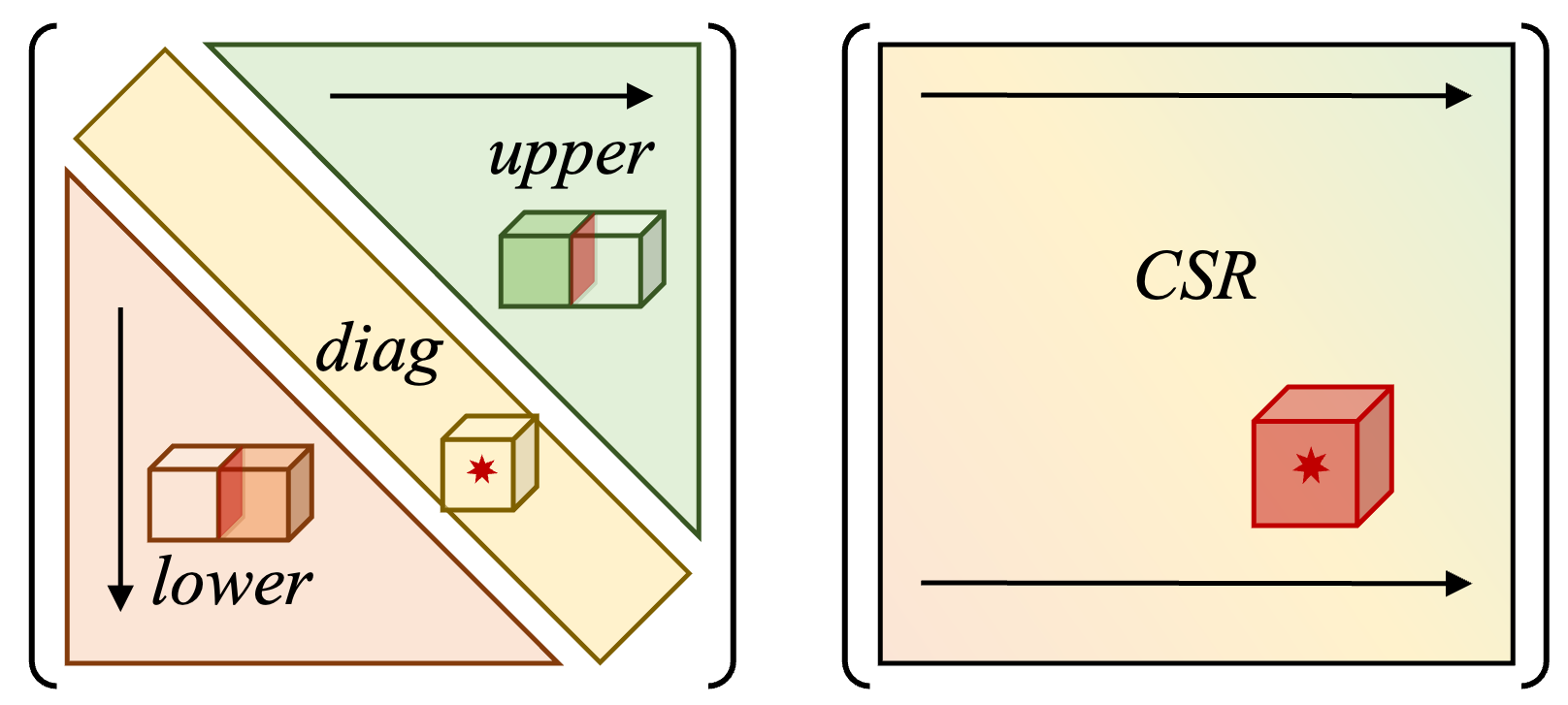}
\caption{Schematic of the {\em ldu} and {\em CSR} formats for sparse matrices. The red color highlights the mesh structure associated with the matrix component, with color intensity indicating the relationship between faces and cells.}
\label{matrix}
\end{figure}

Among these operations, the implementation of implicit solving for PDEs stands out for its high complexity. In contrast to the explicit method, implicit FVM introduces an additional sparse matrix, representing a linear system obtained through discretization. In OpenFOAM, the sparse matrix is stored in {\em ldu} format, where the upper, lower, and diagonal parts are managed individually, as depicted schematically on the left in Fig.~\ref{matrix}. The {\em ldu} format is introduced to ensure good data locality in matrix assembly, taking into account the inherent relationship between the matrix and mesh structure (see Fig.~\ref{matrix}). So it is also employed in our work for matrix assembly.

Providing a more in-depth look into the matrix assembly implementation in our work, the algorithm for discretising the term $\nabla\cdot(\gamma\nabla\psi)$ is outlined in Algo.~\ref{laplacian}. Here, $\gamma$ represents the coefficient of the Laplacian term, and $\psi$ is the unknown field. The subscripts $_f$, $_c$, and $_{bou}$ denote the face, cell, and boundary field, respectively. As seen in line~\ref{algo1:l4} of Algo.~\ref{laplacian}, the upper and lower entries are directly obtained through the multiplication of face values. However, computing the diagonal part necessitates the accumulation of face terms. In the GPU implementation, this poses the challenge of potential simultaneous writes to the same memory location, known as a {\em race condition}. To address this issue, we employ {\em atomicAdd} in lines~\ref{algo1:l5} to \ref{algo1:l6} to calculate the diagonal entries, a method proven to exhibit better GPU performance in 3D FVM compared to the general {\em graph colouring method}.

Handling operations related to boundary fields is detailed in line~\ref{algo1:l8} to \ref{algo1:l12} of Algo.~\ref{laplacian}. Notably, we iterate through the boundary patches and implement distinct kernel functions based on the patch types. This approach accommodates the diverse requirements for various boundary conditions. The multiple kernel launches introduced by the patch loop can be effectively mitigated by leveraging CUDA {\em graph}, as will be introduced in Section~3.2.

While the {\em ldu} format proves suitable for matrix assembly, its performance becomes an issue in {\em sparse matrix-vector products} (SPMVs), the primary operation in solving linear system. This is mainly due to the non-sequential memory access patterns of the off-diagonal matrix parts. Therefore, in our implementation, we convert the matrix from {\em ldu} format to {\em compressed sparse row (CSR)} format (where values are accessed per line, see Fig.~\ref{matrix}), and subsequently utilise the AmgX library to solve the linear system. 

\subsection{Optimisation} \addvspace{10pt}
In this section, we present a series of optimisations to enhance computational performance and minimise the memory footprint on the GPU. Computational efficiency receives a boost through improvements in the kernel function, kernel launch process, and the parallelism of multiple GPUs. Simultaneously, GPU memory usage is curtailed through static data reorganisation and dynamic data allocation approach. 

The initial focus is on the optimisations applied to kernel functions. Essentially, algorithms can be classified into memory-bound and compute-bound categories, with their performance mainly dictated by memory access and elementary computational steps, respectively. In this study, discretization typically display a memory-bound characteristic due to their simple calculations and small stencil sizes. Conversely, operations related to calculating thermophysical and transport properties exhibit a higher arithmetic intensity, categorising them as compute-bound.

For the memory-bound kernels, the paramount optimisation strategy is coalesced data, which serves to augment data locality and reduce data retrieval overhead. In the CPU code, fields undergo operations via cell/face loops, with fields containing multiple components (such as vector fields, tensor fields, and species fields) stored along the component dimension for contiguous data access. However, this data structure disrupts data locality on the GPU. Consequently, we have coalesced the same components of fields to minimise memory access transactions, leading to a significant improvement in bandwidth utilisation efficiency. For the compute-bound kernels, more methodologies are required to enhance the computational performance. Firstly, strategic optimisations have been implemented, featuring computation consolidation and storage substitution. These endeavours alleviate the impact of long-latency computational instructions, particularly the functions such as $sqrt()$ and $pow()$. Moreover, for the intricate computational kernels, 
we have adopted constant memory for storing constant coefficients and shared memory for mass and mole fractions, thereby reducing the dependence on global memory access. Additionally, we have partitioned these heavy kernels into 2 or 3 segments, effectively minimising the register usage of each kernel and achieving higher occupancy. 

The kernel launch procedure is optimised by leveraging CUDA graphs. Conventionally, each execution of a GPU kernel requires a submission operation on the CPU, thereby incurring supplementary overhead. Given the persistent nature of the discretization procedure across each time step, the pertinent kernels can be encapsulated within a CUDA graph. This enables a singular CPU submission to initiate a suite of kernels on the GPU for each time step. This approach markedly augments performance in the discretization procedure. Furthermore, we have opted for the NVIDIA Collective Communication Library (NCCL) over MPI for achieving multi-processor parallelism. Unlike MPI, NCCL supports peer-to-peer communication between GPUs, facilitating direct data exchange between GPUs via PCIe or NVLink. This eliminates the need for CPU-GPU data transfers and further streamlines the communication process. 

\begin{figure}[ht]
\centering
\includegraphics[width=172pt]{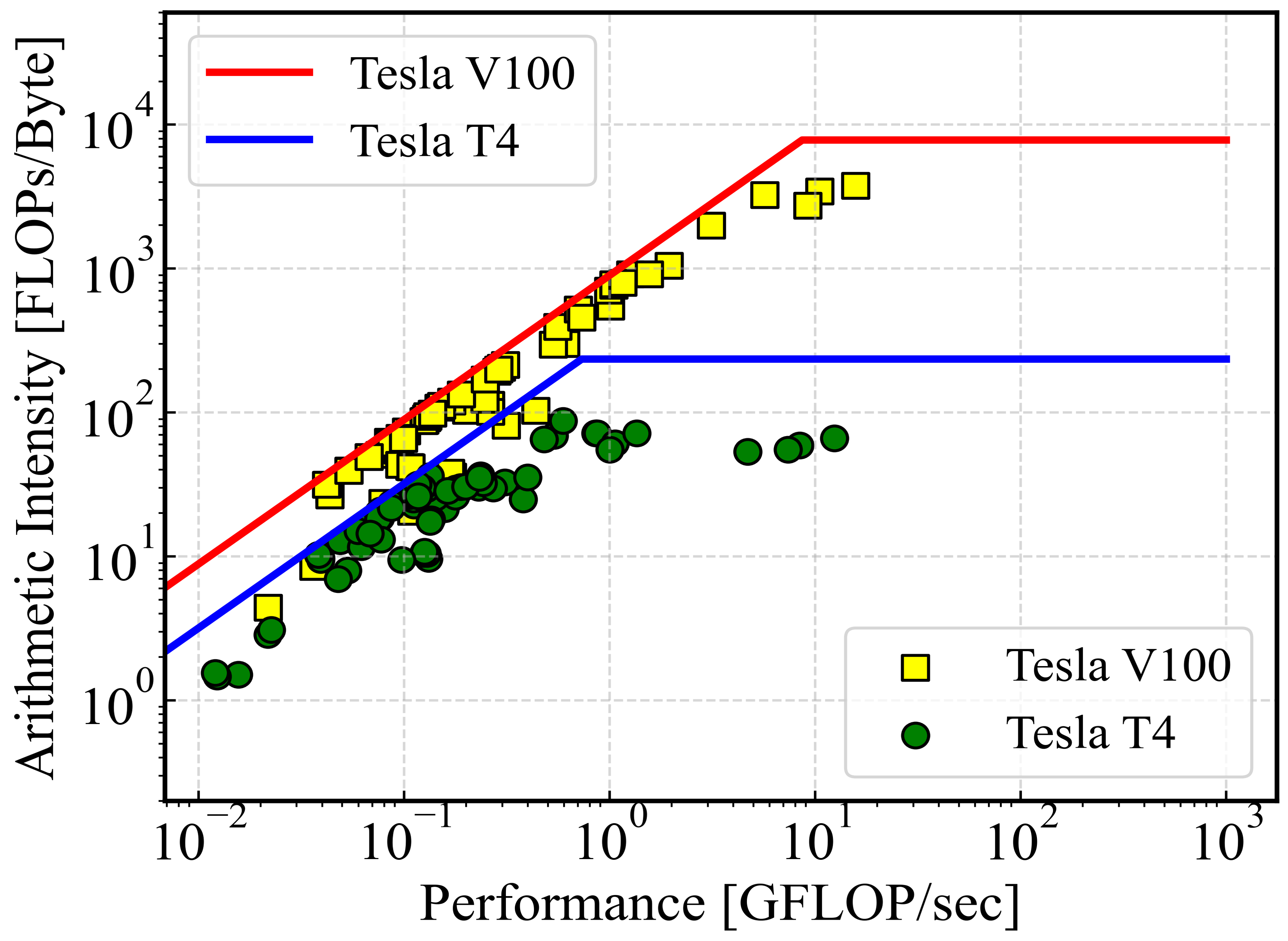}
\caption{Roofline model for V100 and T4 GPUs and mesh size of $128^3$.}
\label{roofline}
\end{figure}

The above optimisations not only yield satisfactory performance for individual kernels but also contribute to the overall simulation efficiency. The performance evaluation is presented here with varying mesh sizes. Notably, with reference to the roofline model illustrated in Fig.~\ref{roofline}, all kernels exhibit commendable efficiency, closely approaching the hardware limit (illustrated by the solid line) on both NVIDIA Tesla V100 and T4 platforms. Furthermore, the overall performance on a single GPU is deemed satisfactory. As illustrated in Table~\ref{singleCardPerf}, the simulation time attains a maximum speedup of approximately 310 at the largest mesh size. Further analysis of performance is expounded upon in Section~4.1.

\begin{table*}[ht] \footnotesize
\centering
\caption{Comparison of computational performance for the H$_2$-air mechanism across various mesh size. Table lists the sum time per step, chemistry time per step, and fluid time per step for a single CPU core and one GPU card.}
\begin{tabular}{cccccccccc}
\hline
\text{Mesh size} & \text{\begin{tabular}[c]{@{}c@{}}CPU \\ sum [s]\end{tabular}}  & \begin{tabular}[c]{@{}c@{}} \text{GPU} \\ \text{sum [s]}\end{tabular}        
    & \text{\begin{tabular}[c]{@{}c@{}}Overall \\ speedup\end{tabular}} & \text{\begin{tabular}[c]{@{}c@{}}CPU \\ chemistry [s]\end{tabular}}  & \text{\begin{tabular}[c]{@{}c@{}}GPU\\ chemistry [s]\end{tabular}}  &
    \text{\begin{tabular}[c]{@{}c@{}}Chemistry\\ speedup\end{tabular}}& \text{\begin{tabular}[c]{@{}c@{}}CPU \\ fluid [s]\end{tabular}}  & \text{\begin{tabular}[c]{@{}c@{}}GPU \\ fluid [s]\end{tabular}}  &
    \text{\begin{tabular}[c]{@{}c@{}}Fluid\\ speedup\end{tabular}}  \\ \hline
$32^3$        & 5.16                                                              & 0.091    & 56.7                                                                      & 4.7   & 0.007    & 671.42  & 0.46   & 0.084    & 5.45           \\
$64^3$          & 37.93                                                                  & 0.183       & 207.26                                                          & 34.24       & 0.053    & 646.03  & 3.69       & 0.13    & 28.38              \\ 
$128^3$           & 281.84                                                         & 0.91       & 309.7                                                                  & 253.86        & 0.41    & 619.1 & 27.938        & 0.5    & 55.876           \\ 
\hline
\end{tabular}
\label{singleCardPerf}
\end{table*}

Finally, we introduce the GPU memory management strategy here. Initially, in comparison to the CPU implementation, we have heightened the granularity of operations, eliminating the storage of numerous intermediate fields in memory. For instance, as delineated in line~\ref{algo1:l3} of Algo.~\ref{laplacian}, the interpolation of cell fields is no longer an independent procedure but has been seamlessly integrated into the Laplacian operation. This strategic consolidation of operations effectively mitigates the burden on GPU memory. Moreover, and of utmost importance, data exclusively utilised in each operation--such as the field of $\nabla U$ for the {\em UEqn} class--is dynamically allocated and freed during each time step. This process is facilitated by the CUDA stream-ordered allocator, which manages memory as stream-ordered operations, and thus circumventing resource-intensive device-wide synchronisation. This approach substantially reduces memory overhead with minimal impact on performance.


\section{Validation and computational performance} \addvspace{10pt}


\subsection{Quasi-DNS of 3D reactive Taylor-Green vortex} \label{sec:QDNS} \addvspace{10pt}

\begin{figure*}[h]
\centering
\includegraphics[width=110mm]{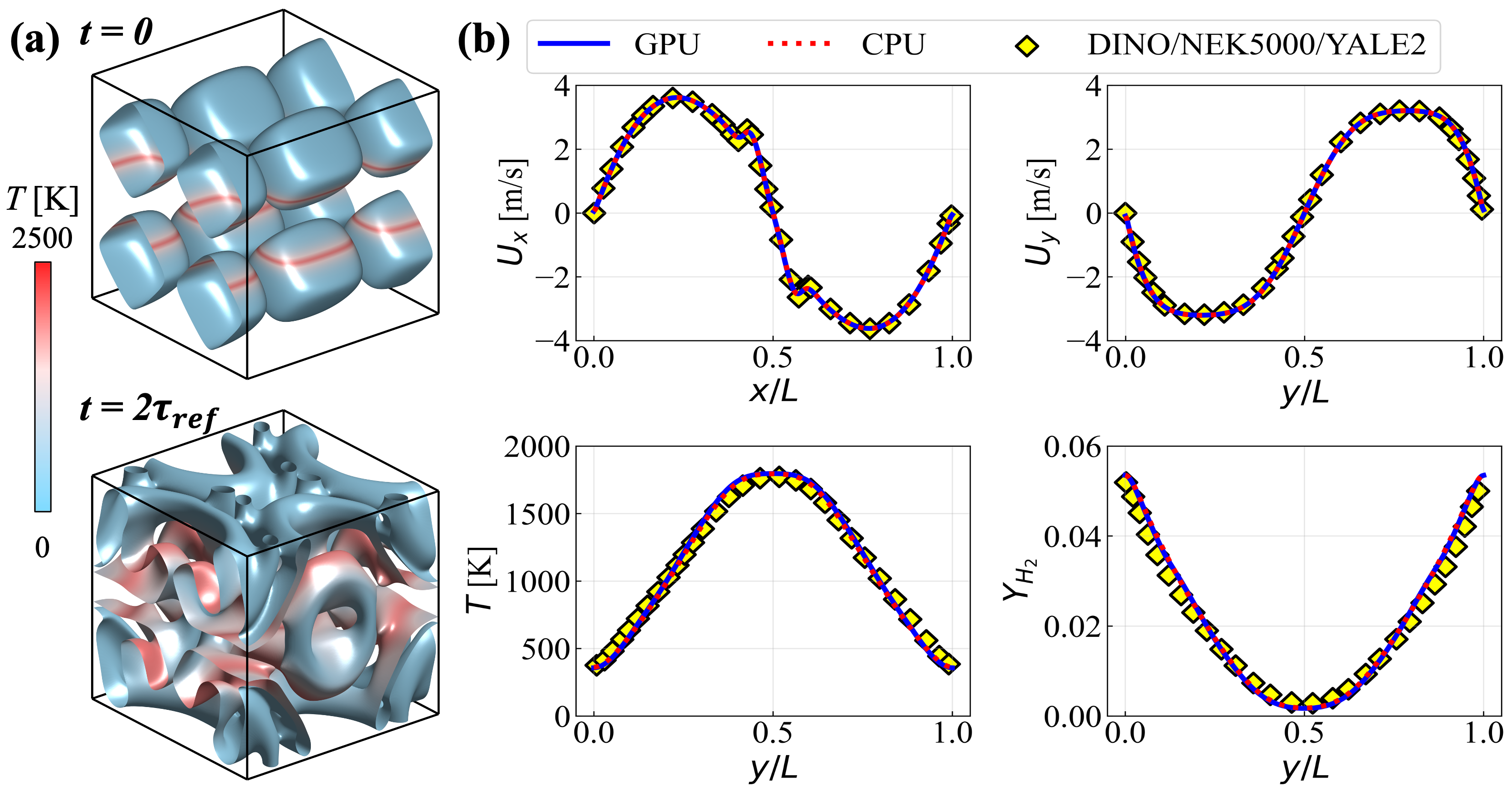}
\caption{(a) The temperature and Q-criterion field at $t = 0$ and $t = 2\tau_{ref}$. (b) Compares of $U_x$, $U_y$, $T$, and $Y_{H_2}$ at $t = 2\tau_{ref}$.}
\label{TGV}
\end{figure*}

As a recently established benchmark \cite{RN4}, the 3D TGV interacting with a H$_2 $ diffusion flame has been widely adopted for code verification and validation \cite{RN111}. The cubic computational domain with an edge length of $2\pi L$ ($L=1$ mm, $Re \approx 250$) is uniformly discretised with $256^3$ cells. The benchmark specified chemical mechanism has 9 species and 12 reversible reactions \cite{boivin2011explicit}. Figure~\ref{TGV}a depicts the TGV evolution of Q-criterion coloured by temperature from the initial field to $t=2\tau_{ref}$ (the flow reference time is $\tau_{ref}=L/u_0$). The initial maximum velocity $u_0=4$ m/s and further details regarding the initial setup can be found in \cite{RN111}. We conducted two simulations of reactive TGV to provide a direct comparison as shown in Fig.~\ref{TGV}b, where GPU indicates full simulation on GPU including the flow and DNN chemistry. In contrast, the CPU simulation follows a standard OpenFOAM procedure with the chemistry solved using CVODE provided by Cantera. As seen, an excellent agreement is obtained between the GPU and CPU simulations, also very close to the reference results from the high-order DNS codes \cite{RN111}. This suggests that our GPU implementation is consistent with the original CPU code and the DNN model coupled with the CFD solver is of good accuracy for complex reactive flow simulation.  

\begin{figure}[h]
\centering
\includegraphics[width=165pt]{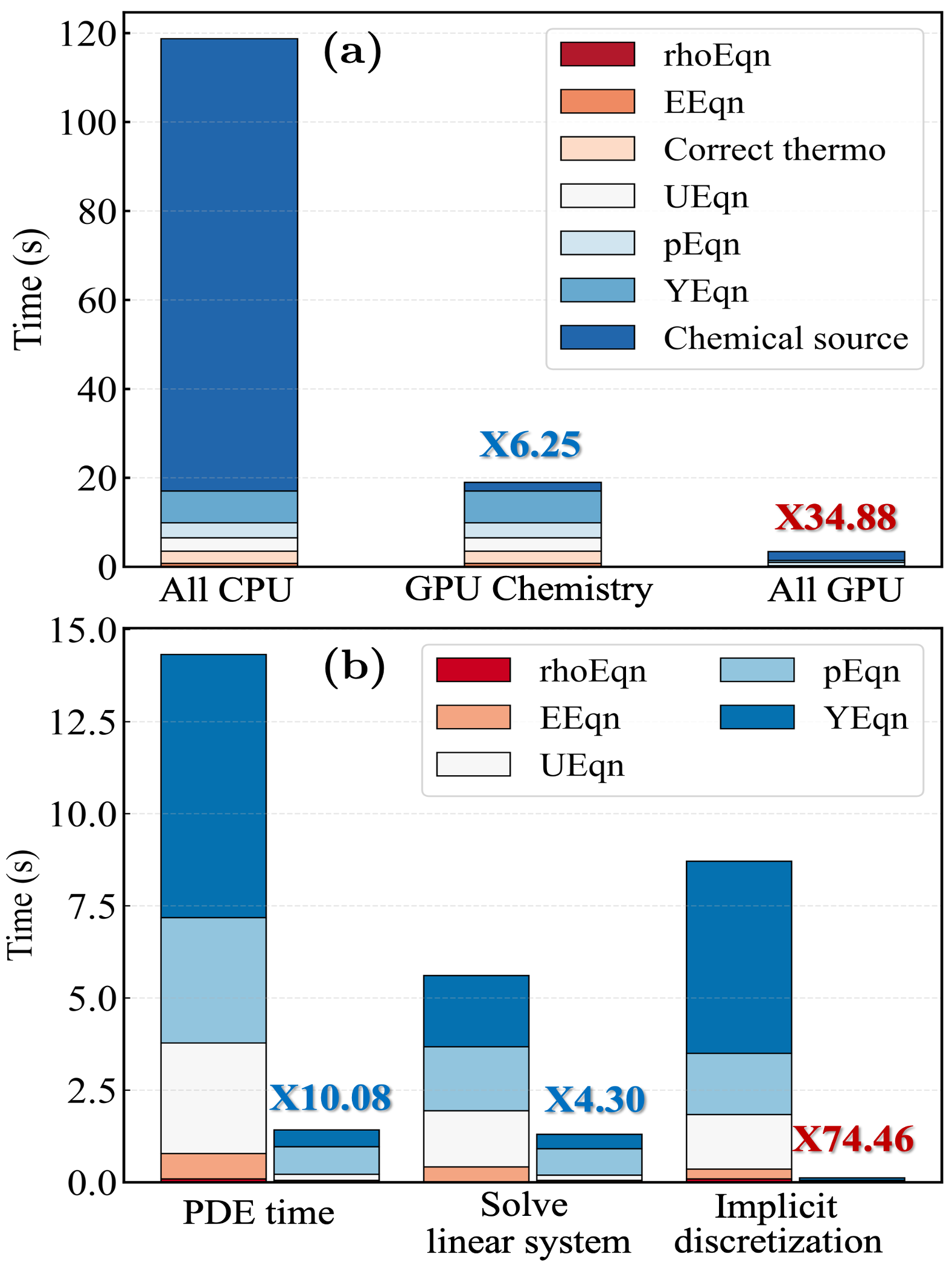}
\caption{Computational cost in simulating 3D reactive TGV in one time step. (a) Overview of the entire simulation process. (b) Breakdown of computational costs specifically for PDE solving components.}
\label{TGVperf}
\end{figure}

The GPU speed-up and memory footprint in the 3D reactive TGV case are analysed and discussed. The CPU simulation is conducted on one CPU chip with 32 processing cores, equivalent to 32 MPI ranks. The GPU results are simulated using 4 NVIDIA V100 GPUs. Note that this configuration is typical for one computing node today (1CPU+4GPU or 2CPU+8GPU). Figure~\ref{TGVperf}a presents the computational cost of each operation outlined in Fig.~\ref{workflow}. Three cases are compared: i) all operations including CVODE chemistry on CPU, ii) flow on CPU with DNN chemistry on GPU, and iii) all on GPU. For the all CPU case, it is as expected that the chemical source integration is the most time-consuming operation. With the excellent speed-up provided by DNN chemistry (demonstrated in Table~\ref{singleCardPerf}), the overall simulation is accelerated by 6.25 times as compared to the all CPU case. After porting all the remaining operations (PDE and thermo/transport) to the GPU, the overall speedup further increases to nearly 35 times. This implies that for a quasi-DNS case with tens of millions of grids, 4 GPU cards within a small workstation at a much lower power cost can provide an equivalent performance to a supercomputer with thousands of CPUs. 

Specifically, Figure~\ref{TGVperf}b details the computational cost for the PDE procedures, which is a key contribution of this work. Bars on the left and right denote CPU and GPU times, respectively. It can be seen that time is evenly distributed in implicit discretisation and solving the linear system when using the CPU. For the GPU solver, however, the operations of discretisation are implemented by CUDA kernels, ultimately achieving a speedup of 75 over the CPU. By contrast, the linear system solving function provided by the AmgX library shows relatively weaker acceleration, especially when solving the pressure Poisson equation. This can mainly be attributed to the fact that the AmgX library cannot access mesh information, limiting it to using the algebraic multigrid method (AMG), whereas the CPU solver adopts the more efficient geometry multigrid method (GAMG). Further optimisation on this implementation is required for better acceleration performance. 

The memory footprint is of particular importance for GPU codes since the memory space is quite limited compared to the RAM size of CPU. For the 4 GPUs used for the TGV case, it is approximately 11 GB on each card and the detailed distribution is listed in Table~\ref{TGVmem}. It is evident that the CFD data, including the mesh and field information, accounts for the majority of the memory, taking up about 44.3\% of the total used. Notably, the memory footprint for storing CFD data has been significantly reduced by 2.27 times with the CUDA stream-ordered allocator. The AmgX handle and CUDA handle together consume 34.86\% and 17.1\% of GPU memory, respectively. These two components primarily consist of compiled kernel functions, CUDA context, and other CUDA resources. Therefore, these portions exhibit minimal growth with the size of the simulation case. The DNN inference procedure utilises minimal memory after optimising both the inference batch size and numerical precision. In summary, with a reasonable GPU memory management strategy, the memory requirement only grows by about 45\% of the memory when the simulation size is doubled (with a larger mesh or chemical mechanism).


\begin{table}[h!] \footnotesize
\caption{The memory usage distribution for each GPU.}
\centerline{\begin{tabular}{ccccc}
\hline 
\text{\begin{tabular}[c]{@{}c@{}}Memory \\ usage \end{tabular}}  & \text{\begin{tabular}[c]{@{}c@{}}CFD \\ data\end{tabular}} 
    & \text{\begin{tabular}[c]{@{}c@{}}AmgX \\ handle\end{tabular}} 
    & \text{\begin{tabular}[c]{@{}c@{}}CUDA \\ handle\end{tabular}} 
    & \text{\begin{tabular}[c]{@{}c@{}}DNN \\ inference\end{tabular}}  \\
\hline
Footprint [MB]& 4978 & 3917 & 1919 & 418  \\
Percentage& 44.33\% & 34.86\% & 17.1\% & 3.72\%  \\ 
\hline 
\end{tabular}}
\label{TGVmem}
\end{table}

\subsection{LES of Cambridge stratified burner} \addvspace{10pt}
The Cambridge burner is a well-known TNF Workshop benchmark for turbulent stratified premixed combustion, previously simulated by many research groups \cite{TNF}. Thus, this configuration is chosen to further demonstrate the practical applicability of the proposed GPU/ML framework. The setup features a central bluff-body surrounded by two co-annular premixed CH$_4$-air mixture streams, with co-flowing air under ambient conditions. The moderately stratified SWB5 case is simulated. The respective velocities of the inner, outer, and air streams are 18.7 m/s, 8.3 m/s, and 0.4 m/s. The stratification is achieved by maintaining an equivalence ratio of 1.0 for the inner and 0.5 for the outer annular flows. 

\begin{figure}[h]
\centering
\includegraphics[width=168pt]{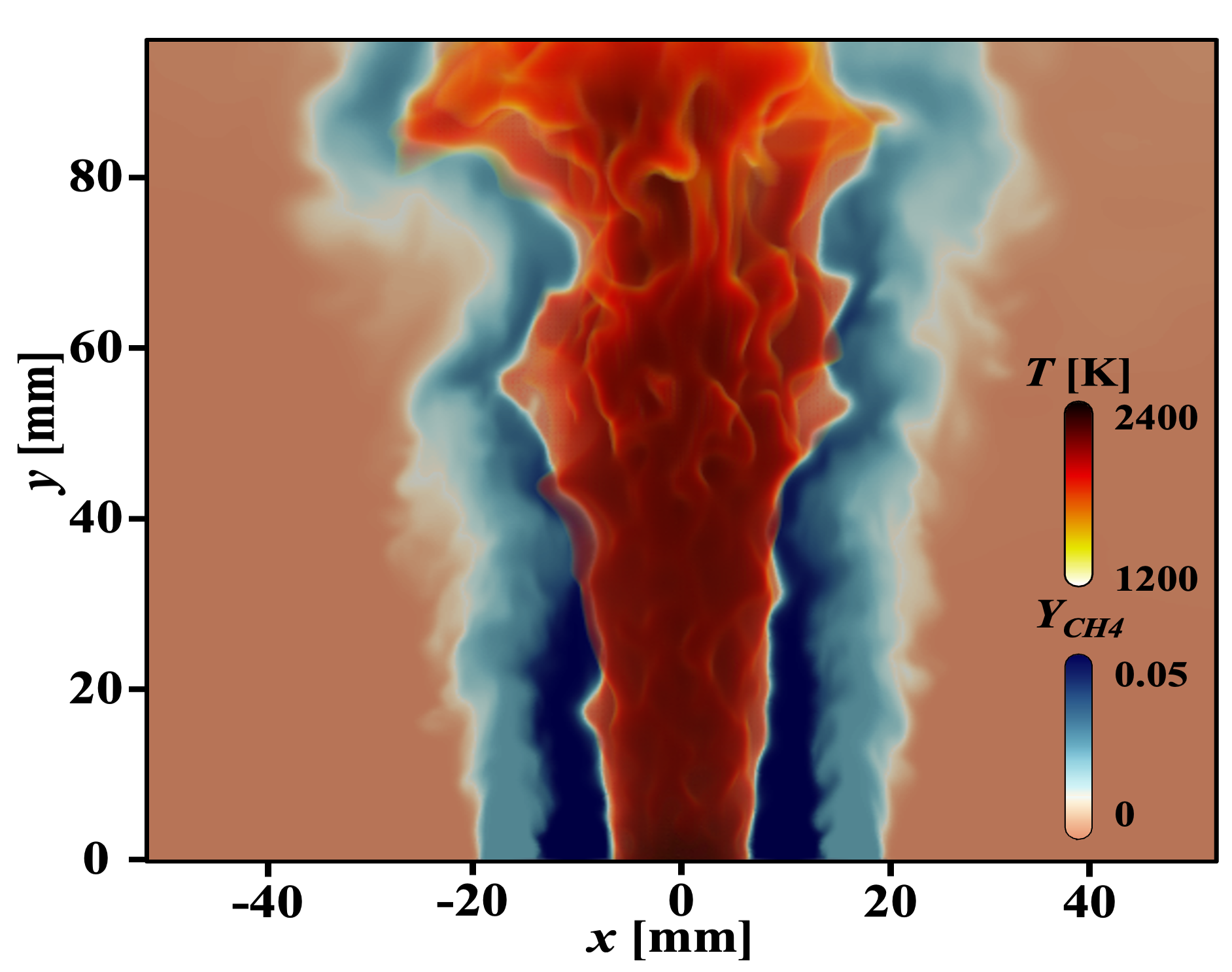}
\caption{Contour plot depicting the mass fraction of methane in the burner mid-section, superimposed with the rendering volume for temperature.}
\label{camflame}
\end{figure}

\begin{figure}[h]
\centering
\includegraphics[width=168pt]{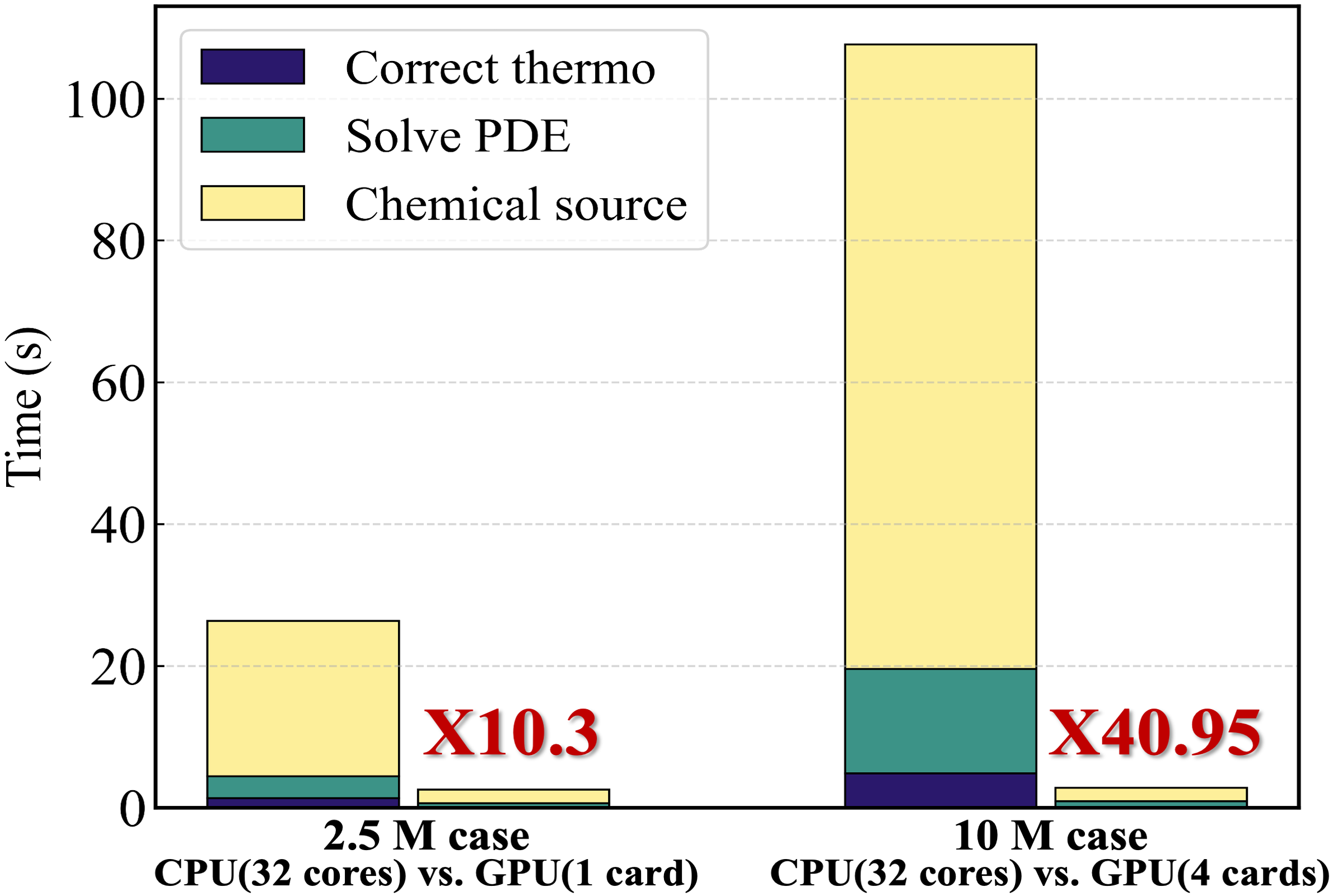}
\caption{One step computational cost in the LES case.}
\label{camflamePerf}
\end{figure}

Figure~\ref{camflame} presents the typical instantaneous temperature and CH$_4$ mass fraction fields obtained from the simulation.
A reduced methane mechanism \cite{drm19} with 20 species and 85 reactions is used. Since the objective here is to assess the computational performance (detailed statistics comparison can be found in our research \cite{zhang2023Cambridge}), two non-uniform grids with of 2.5 and 10 million cells are tested, which covers the typical range for common LES studies \cite{TNF}. 
Time-to-solution comparison is shown for the two cases in Fig.~\ref{camflamePerf}. 
For the 2.5M case, which can be run quite easily using a single GPU, the speedup is 10 times compared to 32 CPU cores. For the 10M case, the simulation is accelerated by a factor of nearly 41 using 4 GPU cards. It is noteworthy that the performance here is improved from that of TGV (see Fig.~\ref{TGVperf}) and this can be attributed to: i) the larger mechanism of CH$_4$ makes the speedup provided by the DNN chemistry more pronounced; ii) the reduced flow complexity (less vortical fields compared to TGV) relaxes the relatively weaker acceleration by AmgX's linear solving, especially for the pressure Poisson equation. However, further implication here is that with the proposed GPU/ML tool, high-fidelity LES of turbulent combustion can be adequately performed on portable devices such as laptops with a consumer-level GPU at the speed of several-hundred core clusters.

\section{Conclusions} \addvspace{10pt}

In this work, we introduce an open-source numerical framework leveraging GPU and ML techniques to accelerate reactive flow simulations. To avoid the CPU-GPU memory copy overhead, the entire computational process is executed on GPU, including ML-accelerated chemistry integration, fully-implicit solving of PDEs, and computation of thermal and transport properties. 
Specifically, the FVM implicit discretisation of conservation equations and explicit computations are implemented through CUDA kernel functions. The multi-processor parallelism is achieved through NCCL, thereby enabling direct communication between GPUs. The linear system is solved using the NVIDIA-produced AmgX library, and ML-related operations are performed with the libTorch library. Various optimisations have been performed to enhance computational performance and reduce GPU memory footprint. 

The capabilities of the framework is evaluated for handling two different turbulent flame test cases using quasi-DNS and LES modelling, respectively. While maintaining similar level of accuracy, for both case the GPU/ML accelerated solver achieves an overall speed-up of over two orders of magnitude. These results underscore the immense potential of machine learning and GPU technologies in empowering reactive flow simulations. In addition, the open-source nature of the proposed framework is expected to facilitate further exploration and collaboration on GPU code development, and bring together diverse machine learning models to become indeed useful for the combustion modelling community.

\acknowledgement{Declaration of competing interest} \addvspace{10pt}


The authors declare that they have no known competing financial interests or personal relationships that could have appeared to influence the work reported in this paper.









 \footnotesize
 \baselineskip 9pt


\bibliographystyle{pci}
\bibliography{PCI_LaTeX}


\newpage

\small
\baselineskip 10pt



\end{document}